\documentclass[amstex,aps,twocolumn]{revtex4}%
\usepackage{graphicx}
\usepackage{color}
\usepackage{bm}
\usepackage{braket}
\usepackage{longtable}
\usepackage{amsmath}
\usepackage{amsfonts}
\usepackage{amssymb}%
 
\begin{document}
\title{Strong and weak single particle nonlocality induced by time-dependent boundary conditions}
\author{S. V. Mousavi }
\affiliation{Department of Physics, University of Qom, Ghadir Blvd., Qom 371614-6611, Iran}
\author{A. Matzkin}
\affiliation{Laboratoire de Physique Th\'{e}orique et Mod\'{e}lisation (CNRS Unit\'{e}
8089), Universit\'{e} de Cergy-Pontoise, 95302 Cergy-Pontoise cedex, France}
\affiliation{Institute for Quantum Studies, Chapman University, Orange, CA 92866, USA}

\begin{abstract}
We investigate the issue of single particle nonlocality in a quantum system
subjected to time-dependent boundary conditions. We first prove that contrary
to earlier claims, there is no strong nonlocality: a quantum state localized
at the center of a well with infinitely high moving walls is not modified by
the wall's motion. We then show the existence of a weak form of nonlocality:
when a quantum state is extended over the well, the wall's motion induces a
current density all over the box instantaneously. We indicate how this current
density can in principle be measured by performing weak measurements of the
particle's momentum.

\end{abstract}
\maketitle

\textbf{Introduction.} Quantum systems with time-dependent boundary conditions
are delicate to handle.\ Even the simplest system -- a particle in a box with
infinitely high but moving walls -- remains the object of ongoing
investigations. From a mathematical standpoint, a consistent and rigorous
framework hinges on unifying an infinite number of Hilbert spaces (one for
each time $t$), each endowed with its own domain of self-adjointness
\cite{dias2011,martino2013,art2015}. The particle in a box with moving walls
has also been taken as a paradigm of quantum chaos, particularly regarding the
existence of Fermi acceleration, ie the unbounded energy gain of a particle
subjected to a time dependent potential
\cite{jose1986,seba1990,leonel2004,glasser2009,lenz2011,grubelnik2014}. More
recently this system has been employed as a tool to investigate expanding
boxes and quantum pistons, particularly in the context of mimicking adiabatic
dynamics without genuine adiabaticity \cite{campo2012}, a technique that is of
interest for the experimental quantum control of different systems, such as
atomic transitions or Bose-Einstein condensates \cite{muga2010}.

The particle in a box with moving walls has also been the prime example in the
investigations of possible nonlocal effects induced by time-dependent boundary
conditions. It was initially suggested by Greenberger \cite{greenberger}, and
subsequently mentioned by several authors, eg
\cite{makowski1992,zou2000,yao2001,wang2008,mousavi2012,mousavi2014}, that
time-dependent boundary conditions could give rise to a genuine form of
nonlocality: a particle at rest and localized in the center of the box,
remaining far from the moving walls, would nevertheless be physically
displaced by the changing boundary conditions induced by the walls motion.

In this paper we distinguish between a strong form and and a weak form of
nonlocality. We first show that there is no strong nonlocality, in the sense
that moving walls have no effect on the dynamics of a \emph{localized quantum
state} placed far from the wall as claimed (but never proved) in previous
works
\cite{greenberger,makowski1992,zou2000,yao2001,wang2008,mousavi2012,mousavi2014}%
. We then show that for quantum states \emph{extended} over the well
extension, the system displays a weak form of nonlocality: moving walls
generate instantaneously a current density in the central region of the box.
We indicate how this nonlocal action could be experimentally tested, namely by
making weak measurements of the particle momentum in the central region of the
box before light has the time to propagate from the walls to that region.
Recent work \cite{braverman2013,steinberg2016} has investigated and
experimentally observed with weak measurements the correlations of the current
density in entangled systems expressed in terms of nonlocal trajectories
assumed to exist in the Bohmian model \cite{holland} of quantum mechanics. In
the present paper, the weak nonlocality we put forward does not rely on
entanglement but is relative to a single particle in a system subjected to
varying boundary conditions.

\textbf{Quantum canonical transformation.} The Hamiltonian for a particle of
mass $m$ in an infinite well of width $L(t)$ with moving boundaries is given
by%
\begin{align}
H &  =\frac{P^{2}}{2m}+V\label{ham}\\
V(x) &  =\left\{
\begin{array}
[c]{l}%
0\text{ \ for}\ \ -\frac{L(t)}{2}\leq x\leq\frac{L(t)}{2}\\
+\infty\text{ \ otherwise}%
\end{array}
\right.  .\label{vdef}%
\end{align}
The solutions of the Schr\"{o}dinger equation $i\hbar\partial_{t}%
\psi(x,t)=H\psi(x,t)$ must obey the boundary conditions $\psi(\pm L(t)/2)=0$.
The instantaneous eigenstates of $H$,
\begin{equation}
\phi_{n}(x,t)=\sqrt{2/L(t)}\cos\left[  \left(  2n+1\right)  \pi x/L(t)\right]
\label{eig}%
\end{equation}
verify $H\ket{\phi_{n}}=E_{n}(t)\ket{\phi_{n}}$ where $E_{n}(t)=\left(
2n+1\right)  ^{2}\hbar^{2}\pi^{2}/2mL^{2}(t)$ are the instantaneous
eigenvalues, but the $\phi_{n}$ are \emph{not} solutions of the
Schr\"{o}dinger equation. Indeed, due to the time-varying boundary conditions,
the problem is ill defined, eg the time derivative $\partial_{t}\psi(x,t)$
involves the difference of two vectors with different boundary conditions
belonging to different Hilbert spaces \cite{martino2013}. To tackle this
problem, different approaches, like introducing a covariant time derivative
\cite{pershogin91} or implementing an ad-hoc change of variables yielding
differential equations involving non-Hermitian operators \cite{makowski91}
have been used. The most general approach to solve the problem is to use a
time-dependent quantum canonical transformation \cite{mosta1999,martino2013}
mapping the Hamiltonian $H$ of the time-dependent boundary conditions to a new
Hamiltonian $\tilde{H}$ of a fixed boundary problem. Let
\begin{equation}
\mathcal{M}(t)=\exp\left(  \frac{i\xi(t)}{2\hbar}\left(  XP+PX\right)
\right)  \label{udef}%
\end{equation}
be a unitary operator with a time-dependent real function $\xi(t)$ defining
the canonical transformation \cite{mosta1999}%
\begin{align}
\ket{ \tilde{\psi}} &  =\mathcal{M}(t)\left\vert \psi\right\rangle
\label{psitrt}\\
\tilde{H}(t) &  =\mathcal{M}(t)H(t)\mathcal{M}^{\mathcal{\dagger}}%
(t)+i\hbar\mathcal{M}(t)\partial_{t}\mathcal{M}^{\mathcal{\dagger}%
}(t)\label{tfh}\\
\tilde{A} &  =\mathcal{M}(t)A\mathcal{M}^{\mathcal{\dagger}}(t)
\end{align}
the latter holding for time-independent observables $A$ such as $X$ or $P$.
Note that $\mathcal{M}(t)$ represents a dilation, ie any arbitrary function
$f(x)$ \ transforms as $\mathcal{M}(t)f(x)=e^{\xi(t)/2}f(e^{\xi(t)}x)$. It is
therefore natural to choose $\xi(t)=\log\left(  L(t)/L_{0}\right)  $ where
$L_{0}\equiv L(t=0)$ so as to map the original problem to the initial interval
$\left[  -L_{0}/2,L_{0}/2\right]  $, with%
\begin{equation}
\psi\left(  x,t\right)  =\left\langle x\right\vert \mathcal{M}%
^{\mathcal{\dagger}}(t)\ket{ \tilde{\psi}}=\sqrt{\frac{L_{0}}{L(t)}}%
\tilde{\psi}\left(  \frac{L_{0}}{L(t)}x,t\right)  .\label{psitr}%
\end{equation}
$\ket{ \tilde{\psi}}$ is the solution of the fixed boundary Hamiltonian
(\ref{tfh}) whose explicit form is%
\begin{equation}
\tilde{H}(t)=\frac{\tilde{P}^{2}}{2m}+V(\tilde{X})-\frac{\partial_{t}%
L(t)}{L(t)}\left(  XP+PX\right)  .\label{hamt}%
\end{equation}

\textbf{Strong nonlocality.} Let us now consider linearly expanding walls,
$L(t)=L_{0}+qt$ with $q>0$. This has been indeed the main case studied in the
context of nonlocality induced by boundary conditions, due to the existence of
exact solutions of the canonically transformed Schr\"{o}dinger equation
$i\hbar\partial_{t}\tilde{\psi}_{n}=\tilde{H}\tilde{\psi}_{n}$.\ By an
educated guess (from the known solutions, originally obtained by inspection
\cite{doescher,makowski91}, of differential equations similar to the
Schr\"{o}dinger equation for $\tilde{H}$) these are found to be given by
\footnote{These solutions are symmetric in $x$, the odd solutions are obtained
similarly but we will not need them here.}%
\begin{align}
\tilde{\psi}_{n}(x,t)=\sqrt{\frac{2}{L_{0}}}  &  e^{\frac{imx^{2}L(t)\left[
\partial_{t}L(t)\right]  }{2\hbar L_{0}^{2}}-i\hbar\pi^{2}(2n+1)^{2}\int
_{0}^{t}L(t^{\prime})^{-2}\,dt^{\prime}/2m}\nonumber\\
&  \cos\left(  \pi(2n+1)x/L_{0}\right)  \label{eigent}%
\end{align}
where $n=0,1,2...$ The $\tilde{\psi}_{n}$ are not eigenfunctions of $\tilde
{H}$, but they can be employed as a fundamental set of solutions in order to
obtain the time-evolved state $\ket{\tilde{\psi}(t)}$ from an arbitrary
initial state $\ket{\tilde{\psi}(t=0)}$ expressed as%
\begin{equation}
\ket{\tilde{\psi}(t)}=\sum_{n}%
\braket{\tilde{\psi_n}(t=0)|\tilde{\psi}(t=0)}\ket{\tilde{\psi_n}(t)}.
\label{td}%
\end{equation}
The solution $\psi(x,t)$ of the original problem with moving boundaries is
recovered from $\tilde{\psi}(x,t)$ through Eq. (\ref{psitr}). In particular,
each solution $\tilde{\psi}_{n}(x,t)$ is mapped into%
\begin{align}
\psi_{n}(x,t)=\sqrt{\frac{2}{L(t)}}  &  e^{\frac{imx^{2}\left[  \partial
_{t}L(t)\right]  }{2\hbar L(t)}-i\hbar\pi^{2}(2n+1)^{2}\int_{0}^{t}%
L(t^{\prime})^{-2}\,dt^{\prime}/2m}\nonumber\\
&  \cos\left(  \pi(2n+1)x/L(t)\right)  . \label{psifund}%
\end{align}

Assume the initial wavefunction is a Gaussian of width $d,$%
\begin{equation}
\left\langle x\right\vert \left.  G(t=0)\right\rangle \equiv G(x,0)=\frac
{(1-i)e^{-\frac{x^{2}}{4d^{2}}}}{2^{3/4}\pi^{1/4}\sqrt{-id}} \label{gau0}%
\end{equation}
with a maximum at the center of the box ($x=0$) and with negligible amplitude
at the box boundaries $x=\pm L_{0}/2$. $\left\vert G(t=0)\right\rangle $ is
expanded over the basis states $\ket{
\tilde{\psi}_{n}(t=0)}$ as per Eq. (\ref{td}) where $g_{n}%
(q)=\braket{\tilde{\psi}_{n}(t=0)|G(t=0)}$ is readily obtained analytically.
The fact that the solutions $\psi_{n}(x,t)$ stretch (in the expanding case) as
time increases has been taken as an indication that the initial Gaussian would
also stretch provided the expansion is done adiabatically so that the
expansion coefficients $g_{n}$ remain unaltered \cite{greenberger}. Hence the
physical state of the particle would be changed nonlocally by the expansion,
although no force is acting on it (we call this strong nonlocality). We show
however that the evolution of the initial Gaussian can be solved exactly in
the linear expanding or retracting cases by using Eqs. (\ref{psitr}) and
(\ref{td}), displaying no dependence of the time-evolved Gaussian on the walls
motion. The periodic case, in which the walls motion reverses and starts
contracting at $T/2$ so that $L(T)=L_{0}$ follows by connecting the solutions
at $t=T/2$.

Our approach to this problem involves the use of special functions, the Jacobi
Theta functions, and a well-known peculiar property of these functions (the
Transformation theorem \cite{bellman}). Let us introduce the Jacobi Theta
function, $\vartheta_{2}(z,\kappa)$, defined here as%
\begin{equation}
\vartheta_{2}(z,\kappa)=2\sum_{n=0}^{\infty}e^{i\pi\kappa\left(  n+1/2\right)
^{2}}\cos\left[  \left(  2n+1\right)  z\right]
\end{equation}
with $\operatorname{Im}(\kappa)>0$. It can be verified that the time evolved
solution $\tilde{\psi}(x,t)=\sum_{n}g_{n}(q)\tilde{\psi}_{n}(x,t)$ can be
summed to yield a Theta function $\vartheta_{2}$, and that further applying
Eq. (\ref{psitr}) gives the wavefunction evolved from $G(x,0)$ as
\begin{equation}
\psi(x,t)=\frac{(1-i)\left(  2\pi\right)  ^{1/4}e^{\frac{imqx^{2}}{2hL(t)}%
}\vartheta_{2}\left(  z,\kappa\right)  }{\sqrt{-idL_{0}L(t)}\sqrt{\frac
{1}{d^{2}}+\frac{2imq}{hL_{0}}}} \label{solTH}%
\end{equation}
with%
\begin{equation}
z=\frac{\pi x}{L(t)};\enskip\kappa=\frac{4\pi\hbar d^{2}}{L_{0}\left(
2d^{2}m\partial_{t}L(t)_{t=0}\right)  -i\hbar L_{0} }-\frac{2\pi\hbar}{m}%
\int_{0}^{t}\frac{1}{L(t^{\prime})^{2}}dt^{\prime} \label{ez}%
\end{equation}
In general $\psi$ as well as $z$ and $\kappa$ depend on $q$, the velocity of
the walls motion. We will explicitly denote this functional dependence, ie
$z(q),\kappa(q).$ The particular case $q=0$ corresponds to static walls with
fixed boundary conditions.

In order to compare the time evolved wavefunction in the static and moving
problems, let us compute $\psi(x,t;q=0)/\psi(x,t;q)$ which after some simple
manipulations takes the form \cite{matzkin-prep}
\begin{equation}
\frac{\psi(x,t;q=0)}{\psi(x,t;q)}=e^{\frac{iz^{2}(0)}{\pi\kappa(0)}%
-\frac{iz^{2}(q)}{\pi\kappa(q)}}\left(  \frac{\kappa(0)}{\kappa(q)}\right)
^{1/2}\frac{\vartheta_{2}\left(  z(0),\kappa(0)\right)  }{\vartheta_{2}\left(
z(q),\kappa(q)\right)  }. \label{rap}%
\end{equation}
We now prove that this expression is unity. The first step is to use the
Jacobi transformation \cite{bellman}
\begin{equation}
\vartheta_{2}\left(  z,\kappa\right)  =\frac{e^{-iz^{2}/\kappa\pi}}{\left(
-i\kappa\right)  ^{1/2}}\vartheta_{4}\left(  \frac{z}{\kappa},-\frac{1}%
{\kappa}\right)  \label{jt}%
\end{equation}
for both $\vartheta_{2}$ functions of Eq. (\ref{rap}). $\vartheta_{4}$ is the
Jacobi Theta function defined by $\vartheta_{4}\left(  z,\kappa\right)
=\sum_{n=-\infty}^{\infty}\left(  -1\right)  ^{n}e^{i\pi\kappa n^{2}}%
e^{2inz}.$ Eq. (\ref{rap}) then becomes%
\begin{equation}
\frac{\psi(x,t;q=0)}{\psi(x,t;q)}=\frac{\vartheta_{4}\left(  \frac
{z(0)}{\kappa(0)},-1/\kappa(0)\right)  }{\vartheta_{4}\left(  \frac
{z(q)}{\kappa(q)},-1/\kappa(q)\right)  }.
\end{equation}

We then note that $\operatorname{Im}-1/\kappa(q)=d^{2}m^{2}L(t)^{2}/\pi\left(
4d^{4}m^{2}+h^{2}t^{2}\right)  $. This is typically a very large quantity:
indeed the typical spatial extension $\Delta x$ of a Gaussian at time $t$ is
deduced from its variance $\left(  \Delta x\right)  ^{2},$ and this quantity
needs to be much less than the spatial extension of the well since by
assumption the particle remains localized at the center of the box, far from
the box boundaries. Hence we have $\left(  \Delta x\right)  ^{2}\ll L_{0}^{2}$
from which it follows that $\operatorname{Im}-1/\kappa(q)\gg1$. Now from the
definition of $\vartheta_{4},$ it is straightforward to see that under these
conditions only the $n=0$ term contributes to the sum, leading to
$\vartheta_{4}\left(  z(q)/\kappa(q),-1/\kappa(q)\right)  =1$ for \emph{any}
value of $q$ (including $q=0$). Hence provided the quantum state remains
localized throughout the evolution, we have
\begin{equation}
\psi(x,t;q)=\psi(x,t;q=0) \label{mr}%
\end{equation}
meaning that the dynamics of the wavefunction initially localized at the
center of the box does not depend on the expanding motion of the walls at the
boundaries of the box.\ In particular the adiabatic condition does not play
any particular role, as Eq. (\ref{mr}) holds for any value of $q$. While each
individual state $\psi_{n}(x,t)$ does stretch out as time increases, the sum
(\ref{td}) for $\psi(x,t)$ ensures that the interferences cancel the
stretching for the localized state. From a physical standpoint there is no
strong nonlocality. The same results holds for walls contracting linearly, as
well as in the periodic case (wall expansion followed by a contraction). In
the latter case it should be noted that the analytic solutions (\ref{eigent})
and (\ref{psifund}) do not verify the Schr\"{o}dinger equation during the
reversal, and as a consequence an expanding basis state $\psi_{n}(x,t)$, does
not evolve into the \textquotedblleft reversed\textquotedblright\ state
$\psi_{n}(x,t)$ after the walls motion reversal \cite{matzkin-prep}.

\begin{figure}[tb]
\includegraphics[angle=-90,origin=c,height=8cm]{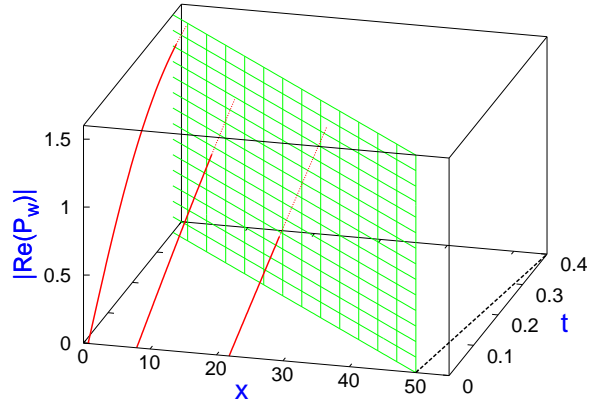}\caption{Weak
nonlocality in an infinite well with expanding walls (only the right half of
the well is shown): each red curve represents $|\operatorname{Re}P_{w}(t)|$
(where $P_{w}$ is the momentum weak value, see Eq. (\ref{Pw})) obtained by
making a weak measurement at the corresponding value of $x$. The wall is
initially at $L_{0}/2=50$ and the green-grided plane represents the boundary
of the light cone originating from $x=50$ at $t=0$. $\operatorname{Re}%
P_{w}(t=0)=0$ but takes significant values before the light cone reaches the
points were $P_{w}$ is computed (the red curves are shown dotted once inside
the light cone). The dashed line to the right schematically represents the
walls motion given by $L(t)$. Numbers given in atomic units, with $L_{0}=100$,
$q=10^{-4}$, $m=1$. }%
\label{figWN}%
\end{figure}

\textbf{Weak nonlocality.} We have seen that the dynamics of an initially
localized state is not modified by moving boundary conditions. However a state
extended all over the box and thus in contact with the walls will naturally be
modified as the boundary conditions change. This process creates a current
density at any arbitrarily chosen point inside the box. For example for
linearly expanding walls, the current density in a basis state $\psi_{n}$ [Eq.
(\ref{psifund})] is
\begin{equation}
j_{n}(x,t)=\frac{2qx\cos^{2}\frac{\pi(2n+1)x}{L_{0}+qt}}{(L_{0}+qt)^{2}}.
\end{equation}
Note that this current density is modified instantaneously, ie $\Delta
j_{n}(x)\equiv j_{n}(x,\varepsilon)-j_{n}(x,0)$ is non-vanishing at $x$ for a
small time interval $\varepsilon$ even if a signal emitted at the wall reaches
$x$ in a time $t_{c}>\varepsilon$ with $t_{c}=\left(  L_{0}/2-\left\vert
x\right\vert \right)  /c$ ($c$ is the light velocity). This is a weak form of
nonocality, in that the modification of the quantum state is due to a local
interaction with the wall but takes place globally and instantaneously in most
of the spatial regions in which the wavefunction amplitude does not vanish.
This weak nonlocality takes a particularly acute form in the de Broglie-Bohm
interpretation \cite{holland} (or Bohmian model, BM): according to BM a
quantum system comprises a point-like particle having an exact but unknown
position, whose motion is guided by the wavefunction. The velocity $v(x,t)$ of
the Bohmian particle is directly linked to the current density%
\begin{equation}
v(x,t)=\frac{j(x,t)}{\left\vert \psi(x,t)\right\vert ^{2}}%
\end{equation}
so that in the BM there is a clear nonlocal action of the boundary motion on
the particle dynamics.

The Bohmian velocity field can be measured by performing weak measurements. A
weak measurement \cite{aav} of an observable $A$ involves a weak coupling
between the system and a pointer followed by a distinct projective measurement
of the system. In the limit of asymptotically weak couplings, the system state
is essentially undisturbed and the pointer is shifted by $\operatorname{Re}%
A_{w}$ where $A_{w}$ is known as the weak value \cite{aav} of $A$. It can be
shown \cite{leavens,wiseman,matzkin2012} that for a system in state
$\left\vert \psi\right\rangle $ the real part of the weak value $P_{w}$ of the
momentum operator $P$ conditioned on a projective measurement at point $x$ is
given by%
\begin{equation}
\operatorname{Re}P_{w}=\operatorname{Re}\frac{\left\langle x\right\vert
P\left\vert \psi\right\rangle }{\left\langle x\right\vert \left.
\psi\right\rangle }=mv(x,t)=\frac{mj(x,t)}{\left\vert \psi(x,t)\right\vert
^{2}}.\label{Pw}%
\end{equation}
Weak momentum values have already been experimentally determined for photons
\cite{kocsis} and specific proposals to perform such measurements with single
electron sources have been put forward very recently \cite{oriols}.

Fig. 1 displays $\operatorname{Re}P_{w}(t)$ when the system is initially in
the quantum state $(\phi_{10}(x,0)-\phi_{1}(x,0))/\sqrt{2}$ where $\phi
_{1}(x,0)$ given by Eq. (\ref{eig}) is an eigenstate of the \emph{fixed} walls
box, with $j(x,0)=0$. We allow for a continuous transition from the fixed
walls to the linear regime by setting $L(t)=L_{0}+qt(1-e^{-\beta t})$ with
$\beta\gg1$. The light cone boundary is indicated by the green-grided plane.
It can be seen that the current density reacts to the walls motion before a
signal can reach the point where the weak measurement takes place. This is the
signature of weak nonlocality. Whether such an effect can be experimentally
observed in practice (under the present configuration, an experiment would
require to carry out a weak measurement in a sub-femtosecond timescale) as
well as its status relative to the no signaling principle remains to be
investigated. Note that the nature of single particle nonlocality here is
different than in the case of entangled particles, for which weak momentum
measurements have been recently performed \cite{steinberg2016, braverman2013}
by establishing correlations between the polarization of one photon and the
current density of the other.

\textbf{Conclusion.} To sum up we have shown that contrary to widespread
claims, time-dependent boundary conditions do not induce a strong form of
nonlocality that would modify the dynamics of a quantum state entirely
localized at the center of a box. However when the state of the system is
extended over the box a weaker form of nonlocality is induced by the varying
boundary conditions: a current density appears instantaneously at any point of
the box, however far from the moving walls. This effect can in principle be
tested experimentally by performing weak measurements.

\emph{Acknowledgements}\textbf{.} AM would like to thank Dipankar Home (Bose
Institute) for early discussions on the issue of strong nonlocality in an
expanding well and the participants of a FQXI workshop in Marseille (July
2017) for useful exchanges on weak nonlocality.

\end{document}